\documentclass[twocolumn,showpacs,showkeys,pre,floatfix]{revtex4}
\usepackage{bm}
\usepackage{graphicx}
\usepackage{color}

\newcommand{\singlefiguresize}{0.80\columnwidth}
\newcommand{\doublefiguresize}{1.90\columnwidth}

\begin{document}

\title{Design of Easily Synchronizable Oscillator Networks Using the Monte
Carlo Optimization Method}
\author{Tatsuo Yanagita}
\affiliation{Research Institute for Electronic Science, Hokkaido University, Sapporo
001-0020, Japan}
\email{yanagita@nsc.es.hokudai.ac.jp}
\homepage{http://www-nsc.es.hokudai.ac.jp/~yanagita}
\author{Alexander S. Mikhailov}
\affiliation{Fritz-Haber-Institut der Max-Planck-Gesellschaft, Faradayweg 4-6, 14195
Berlin, Germany}

\begin{abstract}
Starting with an initial random network of oscillators with a heterogeneous frequency
distribution, its autonomous synchronization ability can be largely improved by appropriately rewiring the links between the elements. Ensembles of
synchronization-optimized networks with different connectivities are
generated and their statistical properties are studied.
\end{abstract}

\date{\today }
\maketitle



\section{Introduction}

In the last decade, much interest has been attracted to studies of complex
networks consisting of dynamical elements involved in a
set of interactions \cite{RevModPhys.74.47,Boccaletti2006175}. Particular
attention has been paid to problems of synchronization in
network-organized oscillator systems \cite{Manrubia04,Arenas08}. Synchronization
phenomena are ubiquous in various fields of science and play an important
role in the functioning of living systems \cite{Kurths01}.
Investigations focused on understanding the relationship between the
topological structure of a network and its collective synchronous behavior 
\cite{Boccaletti2006175}. Recently, synchronization properties of systems
formed by phase oscillators on static complex networks, such as small-world
networks \cite{Hong02} and scale-free networks \cite{Ichinomiya04,Lee05},
have been considered. It has also been shown that the ability of a network
to give rise to synchronous behavior can be greatly enhanced by exploiting
the topological structure emerging from the growth processes \cite{PhysRevE.71.016116,PhysRevLett.94.138701}. However, full understanding of
how the network topology affects synchronization of specific dynamical units
is still an open problem.

One possible approach is to use evolutionary learning mechanisms in order to
construct networks with prescribed dynamical properties. Several models have
been explored, where dynamical parameters were modified in response to the
selection pressure via learning algorithms, in such a way that the system
evolved towards a specified goal \cite{Mikhailov02,Moyano01,Gleiser06,PhysRevE.80.066120,Brede20085305,Brede08}. In our study,
this approach is employed to design phase oscillator networks with
synchronization properties. We consider adaptive evolution of a
network of coupled heterogeneous phase oscillators \cite%
{Kuramoto84,RevModPhys77137}. In such systems, heterogeneity of
oscillator frequencies competes with the coupling which favors emergence
of coherent dynamics \cite{Kuramoto84,Manrubia04}. The question is how to
connect a set of phase oscillators with given natural frequencies, so
that the resulting network would exhibit the strongest synchronization,
under the constraint that the total number of available links is fixed.

Previously, a related, but different problem of synchronization optimization
in a network with the fixed topology through the modification of connection
weigths was considered \cite{tanaka08}. Assuming that the system was in
a phase-locked state, the deterministic steepest descent method was used to determine the coupling strengths between elements which lead to the
best possible phase synchronization. In contrast, we consider the systems
which stay in partially synchronized states (that is, are not fully
phase-locked) and ask what should be the optimal topology of connections,
with each link having the same strength. 

To design optimal networks, stochastic Markov Chain Monte Carlo (MCMC) method with replica exchange is
used by us. Large ensembles of optimal networks are constructed and their
common statistical properties are analyzed. As we observe, the typical
structure of a synchronization-optimized network is strongly dependent on
its prescribed connectivity. Sparse optimal networks, with a small number of
links, tend to display a structure with relatively high clustering, similar
to that found for the networks of chaotic maps \cite{Gong04,Berg04}. As the
connectivity is increased, synchronization-optimized networks show a
transition to (approximately) bipartite architectures.

The paper is organized as follows. In Sec.~\ref{sec:model}, we introduce a
model of heterogeneous phase oscillators occupying nodes of a directionally
coupled network and define the synchonization measure for this system. The
optimization method is also introduced in this section. 
Construction of the optimized networks and their statistical analysis are performed in Sec.~\ref{sec:numerical}. The results are finally discussed in Sec.~\ref{sec:summary}



\section{The Model and the Optimization Method}

\label{sec:model}

We consider $N$ oscillators with different natural frequencies placed onto
the nodes of a network. The evolution of this system is given by 
\begin{equation}
\frac{d\theta _{i}}{dt}=\omega _{i}+\frac{\lambda }{N}\sum_{j=1}^{N}a_{i,j}%
\sin (\theta _{j}-\theta _{i}),  \label{eq:model}
\end{equation}%
where $\omega _{i}$ is the natural frequency of oscillator $i$ and $\lambda $
is the coupling strength. The weights $a_{i,j}$ define the adjacency matrix $%
\mathbf{a}$ of the interaction network: $a_{i,j}=1$ if oscillator $i$
interacts with oscillator $j$, and $a_{i,j}=0$ otherwise. The adjacency
matrix is generally asymmetric.

To quantify synchronization of the oscillators, the Kuramoto order parameter 
\begin{equation}
r(t)=\frac{1}{N}\left\vert \sum_{i=1}^{N}\exp (i\theta _{i})\right\vert
\end{equation}%
is employed. Under perfect synchronization, we have $r=1$, whereas $r\sim 
\mathcal{O}(N^{-1/2})$ in absence of coupling for randomly drawn natural
frequencies. A second-order transition takes place at some critical coupling
strength $\lambda _{c}$ from the desynchronized to the synchronized states \cite{Kuramoto84}. 

To measure the degree of synchronization, we numerically integrate Eq.~(\ref%
{eq:model}) for given initial conditions $\theta _{i}(t=0)\in \lbrack 0,2\pi
)$ and calculate the average modulus of $r(t)$ over a long time $T,$ 
\begin{equation}
R(\mathbf{a})=\left\langle \frac{1}{T}\int_{0}^{T}r(t)dt\right\rangle
_{init.},
\end{equation}%
where $\langle \dots \rangle _{init.}$ represents an average over many
realizations with different initial conditions $\theta _{i}(0)$.

Our aim is to determine the network $\mathbf{a}$ which would exhibit the
highest degree of synchronization, provided that the total number of links
is fixed and a set of natural frequencies is given. The network construction
can be seen as an optimization problem. The optimization task is to maximize
the order parameter and, possibly, bring it to unity by changing the network 
$\mathbf{a}$. An approximate standard approach to the problems of complex
combinatorial optimization, such as the traveling salesman problem, is
provided by the method of simulated annealing (see, e.g. \cite{Mikhailov02}%
). However, we are interested in the \textit{statistical properties} of the
synchronization-optimized networks rather than in a search for the
best-optimized network. If multiple samples are generated using conventional
optimization methods such as simulated annealing, it is difficult to control
the probability of the repeated appearance of the same (or similar) items in
the obtained set of samples.

To study statistical ensembles of optimized networks, the MCMC method \cite{Landau05,Newman99,Liu01},
which has previously been applied to dynamical systems \cite{Cho94,Bolhuis98,Vlugt00,
Kawasaki05,Sasa06,Giardin06,Tailleur07,yanagita09},
will be used. The canonical ensemble average of a
network function $f(\cdot )$ is introduced as  
\begin{equation}
\overline{f_{\beta }}=\sum_{w}\frac{f(\mathbf{a})\exp (\beta R(\mathbf{a}))}{%
Z(\beta )},  \label{eq:gibbs}
\end{equation}%
where $Z(\beta )=\sum_{w}\exp (\beta R(\mathbf{a}))$ is the partition
function and the parameter $\beta $ plays the role of the inverse
temperature. 

Hence, the problem is reduced to sampling from the ensemble with the Gibbs
distribution $\exp (\beta R(\mathbf{a}))$. Such ensemble can be generated,
for example, by using the Metropolis algorithm~\cite{Metropolis53}, which is
the simplest implementation of the MCMC method. The Metropolis algorithm,
which we use, is essentially standard. The only important difference is that
we should simulate the dynamics with a network $\mathbf{a}$ at each iterated
trial. 

This Metropolis algorithm appears to provide a simple and universal way of
generating the Gibbs network distribution. However, the efficiency of such
algorithm gets worse when $\beta $ increases, particularly in the case of a
highly jagged landscape $R(\mathbf{a})$. This deficiency can be eliminated
by using instead the Replica Exchange Monte Carlo (REMC) algorithm, which
provides an efficient method to investigate systems with rugged free-energy
landscapes, specifically at low temperatures \cite{Hukushima96,Iba01,Janke08}%
.

In a REMC simulation, a number of replicas $\left\{ \mathbf{a}_{m}\right\}$ %
with different inverse temperatures $\beta _{m\text{ }}$are evolved in
parallel. At regular evolution time intervals, the performances of a
randomly selected, adjacent pair of replicas are compared.  The running
configurations of the two selected replicas are exchanged with the
probability $\min \left[ 1,\exp \left( \Delta \beta \Delta R\right) \right] $%
, where $\Delta \beta =\beta _{m+1}-\beta _{m}$ is the difference of the
inverse temperatures of the pair and $\Delta R=R(\mathbf{a}_{m+1})-R(\mathbf{%
w}_{m})$ is the difference of their performances. The exchange of replicas
with different temperatures effectively imitates repeated heating and
annealing, thus preventing trapping in the local performance optima. Note
that such stochastic exchange algorithm preserves the joint probability
distribution $\Pi _{m}\exp \left( \beta _{m}R(\mathbf{a}_{m}\right)
)/Z(\beta _{m})$, so that the unbiased set of samples is generated for all
inverse temperatures.

Explicitly, the algorithm is defined as follows:

\begin{enumerate}
\item The states of replicas $\{ \mathbf{a}^{0}_{m} \}$  are initialized by random networks 
(which is chosen as a random Erd\"{o}s -R\'{e}nyi network)

\item The candidate for the next network $\mathbf{a}^{\prime}_{m}$ at iteration
step $n$ is obtained from the current network $\mathbf{a}^{(n)}_{m}$ by rewiring
one of its links. A randomly chosen link is moved to a randomly chosen link
vacancy, so that the total number of links remains conserved.

\item The evolution equations~(\ref{eq:model}) for the network $\mathbf{a}_{m}%
^{\prime }$ are integrated using the standard Euler algorithm. The order
parameter is then calculated and averaged over the time interval $t\in
\lbrack 0,T]$ and over a fixed number of realizations starting from
different random initial conditions. Thus, the synchronization property $R(%
\mathbf{a}^{\prime }_{m})$ of the candidate network is determined.

\item Next, a random number $x\in \lbrack 0,1]$ is uniformly drawn. If  
\[
x<\frac{\exp (\beta R(\mathbf{a}^{\prime }_{m}))}{\exp (\beta R(\mathbf{a}%
^{(n)}_{m}))}\ ,
\]%
the candidate is accepted and taken as  $\mathbf{a}^{(n+1)}_{m}=\mathbf{a}%
^{\prime }_{m}$; otherwise nothing is changed, so that $\mathbf{a}^{(n+1)}=%
\mathbf{a}^{(n)}_{m}$.

\item At regular evolution time intervals, the performances of a
randomly selected, adjacent pair of replicas are compared.  The running
configurations of the two selected replicas are exchanged with the
probability 
\[
\min \left[ 1,\exp \left\{ (\beta _{m+1}-\beta _{m}) (R(\mathbf{a}^{(n+1)}_{m+1})-R(\mathbf{
w}^{(n+1)}_{m}) \right\} \right].
\]

\item  Return to Step~(2) until the statistical average Eq.~(\ref{eq:gibbs}) converges.

\end{enumerate}

\begin{figure}[tbp]
\begin{center}
\resizebox{\singlefiguresize}{!}{\includegraphics{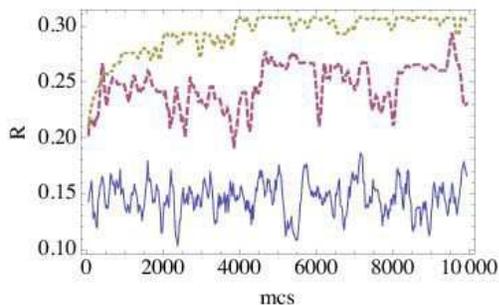} }
\end{center}
\caption{(Color online). The evolution of order parameters of coupled
oscillator networks during the optimization process. The blue solid, red broken, and
yellow dotted lines are for $\protect\beta =\protect\beta _{0},\protect\beta _{5}$
and $\protect\beta _{M}$, respectively. Note that the blue solid line is for $%
\protect\beta _{0}=0$ and, therefore, it corresponds to the networks
generated by only random rewiring. The parameters are $p=0.2,N=20,\protect%
\lambda =1.0,\protect\gamma =0.3,M=21,\protect\delta \protect\beta =10$.}
\label{fig:timeseries}
\end{figure}

\section{Numerical analysis}
\label{sec:numerical}
To determine the synchronization degree of a given network at each iteration
step of the optimization procedure, equations~(\ref{eq:model}) were
numerically integrated with the time increment $\Delta t=0.05.$ Averaging
over five independent realizations started from different random initial
conditions has been furthermore performed at each iteration step. Oscillator
ensembles of sizes $N=10$ and $20$ were considered. Natural frequencies of
the oscillators were always chosen as $\omega _{i}=-\gamma +2\gamma i/N$ ,
so that they uniformly distributed within the interval $[-\gamma ,\gamma
]$ \footnote{%
We have also performed simulations with randomly selected natural
frequencies, which were drawn from a distribution $g(\omega )$. The
qualitative behavior was similar to that of the present model.}. 

Initial phases $\theta _{i}(0)=2\pi f_{init}(i)/N$ uniformly distributed
inside the interval $[0,2\pi )$, where $f_{init}(i)$ is a random one-to-one
mapping between $\{1,\cdots ,N\}$. Hence, the order parameter at $t=0$
always zero. To construct initial random networks with a given number $K$ of
connections and, thus, the connectivity $p=K/N(N-1)$, $K$ off-diagonal
elements of the adjacency matrix were randomly and independently selected
and set equal to unity.

For time averaging, intervals of length $T=100$ and $200$ were typically
used. The results did not significantly depend on $T$ when sufficiently
large lengths $T$ were taken. Using the order parameter, graphs $\mathbf{a}$
were sampled by the REMC optimization method. In parallel, evolution of $M$
replicas with the inverse temperatures $\beta _{m}=\delta \beta \times m,\
m=0,1,\dots ,M$ was performed (with $M=21$ and $\delta \beta =10$). At each
five Monte Carlo steps (mcs), the perfomances of a randomly chosen pair of
replicas were compared and exchanged, as described above. For display and
statistical analysis, sampling at each every 50 mcs after a transient of $%
5000$ mcs has been undertaken.

\begin{figure*}[tbp]
\begin{center}
\resizebox{\doublefiguresize}{!}{\includegraphics{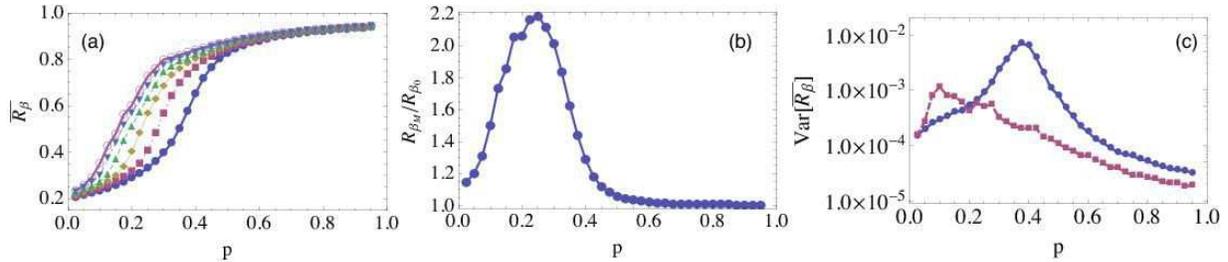} }
\end{center}
\caption{(Color online). Average order parameters  as functions of the
network connectivity $p$. The blue filled circles are for the replica  $\protect%
\beta _{0}$, i.e., the ensemble of  randomly rewired networks. The red squares,
yellow diamonds, green triangles, blue inverted triangles, and red open circles are for replicas
with $\protect\beta =40,80,120,160,$ and 200, respectively. 
(b) Ratio of the average order parameters for the
synchronization-optimized ensemble with the inverse temperature $\protect%
\beta _{M}$ and for $\protect\beta _{0}=0$. 
(c) Variance of the order parameters. The red squares are for the random rewired
ensemble, the blue circles are for the synchronization-optimized ensemble. The
same parameters as in Fig.~\protect\ref{fig:timeseries}}
\label{fig:order}
\end{figure*}

\subsection{Optimization at different temperatures}

Synchronization-optimized networks were obtained by running the evolutionary
optimization. In this process, the order parameter was progressively
increasing until a stationary state has been achieved. Figure~\ref%
{fig:timeseries} displays the optimization processes at different
temperatures. As clearly seen, when using replicas with the larger inverse
temperature $\beta ,$ the larger values of the order parameter could be
reached, although the optimization process was then more slow. 
After the transients, statistical averaging of the order parameter over the ensemble
with the Gibbs distribution has been performed, according to Eq.~(\ref%
{eq:gibbs}).

In Fig. ~\ref{fig:order}(a), the averaged order parameter $\overline{R}$ is
displayed as a function of the connectivity $p$ for several different
inverse temperature $\beta $. The blue solid circle symbols  show the averaged
order parameter corresponding to the replica with $\beta _{0}=0,$ i.e. for
an infinitely high temperature. We see that the averaged order parameter
increases with the network connectivity $p$ even if the networks are
produced by only random rewiring. The red open circles show the average order
parameters for the ensemble corresponding to the replicas with the lowest
inverse temperature $\beta _{M}$. Generally, greater order parameters are
obtained by running evolution at higher inverse temperatures  $\beta $ at
any network connectivity $p$. At each connectivity $p$, the order parameter
is gradually increased with increasing $\beta $ and is approximately
saturated at $\beta _{M}$. This means that, even if one further increases $%
\beta $, only slight improvements of the averaged order parameter can be
expected. Thus, the networks sampled by the replica with the largest inverse
temperature $\beta _{M}$ are already yielding a representative optimal
ensemble. 

Figure~\ref{fig:order}(b) shows, depending on the network connectivity $p$, 
the ratio $\overline{R_{\beta _{M}}}/\overline{R_{\beta _{0}}}$ of the
averaged order parameters sampled by the optimal network ensemble with  $%
\beta _{M}$ to those obtained for the ensemble with purely random rewiring.
Since there is no room for the improvement of the order parameter when the
number of links is small, the ratio tends to unity as the connectivity  $p$
is decreased. On the other hand,  when $p=1$, global coupling is realized,
for which, under the chosen coupling strength, full synchronization occurs.
As evidenced by this figure, the difference between the synchronization
capacities of the optimzed and random networks is most pronounced at the
intermediate connectivities $p$.

In Fig.~\ref{fig:order}(c),  the mean variance  $\overline{\mbox{Var}\lbrack
R]_{\beta }}=\overline{R_{\beta }^{2}}-\overline{R_{\beta }}^{2}$ of the
order parameters at different connectivities $p$ is displayed. It can be
observed that this mean variance for the synchronization-optimized ensemble
decreases with an increase in the number of links, while the respective mean
variance for the  random rewired ensemble has a maximum at $p=0.4$. Note
that, since the transition from the connected to the disconnected random
graphs occurs at  $p_{c}=1/N$ \cite{RevModPhys.74.47,Boccaletti2006175},
this behavior is not directly related to the topological transition in the
network itself.
\begin{figure*}[tbp]
\begin{center}
\resizebox{\doublefiguresize}{!}{\includegraphics{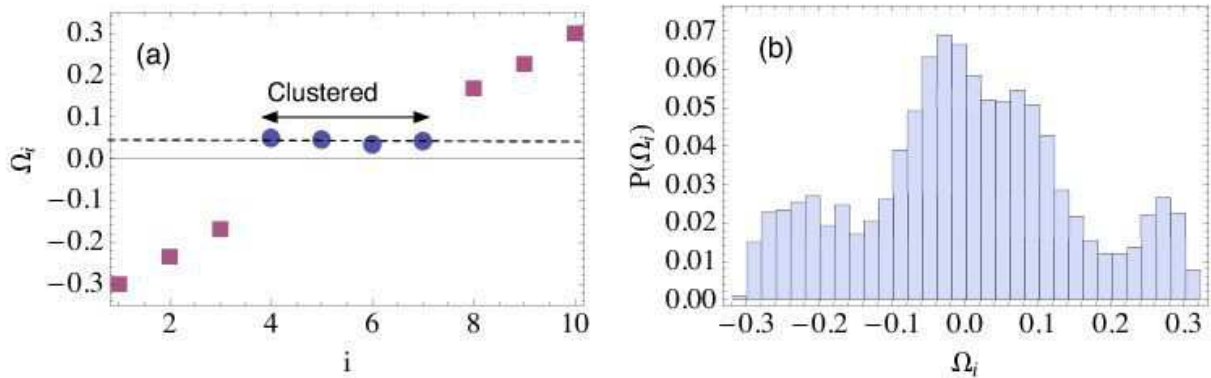} }
\end{center}
\caption{(Color online). (a) Time-averaged frequencies (winding numbers) of
the oscillators in a synchronization-optimized network. (b) Statistical
distribution of winding numbers for the synchronization-optimized
ensemble.The parameters are $p=0.2$, $\protect\beta =\protect\beta _{M}$;
other are the same as  in Fig.~\protect\ref{fig:timeseries}. }
\label{fig:winding}
\end{figure*}

To further analyze the behavior of oscillators in synchronization-optimized
networks, we calculated time-averaged frequencies, i.e., winding numbers $%
\Omega _{i}=(1/T)\int_{0}^{T}\theta _{i}(t)dt$ of all oscillators $i.$
Histograms of distributions over the winding numbers were constructed by
counting the numbers of oscillators with the winding number inside a fixed
bin interval, $H_{k}=\{\Omega _{i}|n\delta \Omega <\Omega _{i}<(k+1)\delta
\Omega \}$, where $k=0,1,\dots ,K-1$, $K=10$ is the number of bins, and $%
\delta \Omega =2\gamma /K$ is the bin size. The winding number as a function
of the natural frequency is shown in Fig.~\ref{fig:winding}(a). The blue circles
show the entrained cluster with the winding number approximately equal to
zero. The cluster consists of the elements whose natural frequencies are
near the mean natural frequency $\Omega =0$. While the specific elements of
the cluster and its size depend on a particular network in the
synchronization-optimized ensemble, there is a statistical trend that the
entrained cluster consists of the oscillators in the neighborhood of the
zero frequency. This is demonstrated by the histogram of winding numbers for
the synchronization-optimized ensemble in Fig.~\ref{fig:winding}(b). Note
that the oscillators are always ordered according to their natural
frequencies $\omega _{i}=-\gamma +2\gamma i/N$ which monotonously increase
with $i$. We see that all elements get divided into two groups, in which $%
\Omega _{i}\approx 0$ or where the winding number is relatively high. For each
particular network realization, there should be a peak at the frequency of
the entrained cluster. The position of this peak depends however on the
realization and, as a result, the histogram of the winding numbers for the
entire ensemble shows a broad maximum. This behavior is characteristic for
relatively low connectivities. The broad peak gradually sharpens when the
connectivity is increased because the size of the cluster increases and
fluctuations of the winding number become smaller.

\subsection{Architectures of Synchronization-Optimized Networks}

\begin{figure}[tbp]
\begin{center}
\resizebox{0.9\columnwidth}{!}{\includegraphics{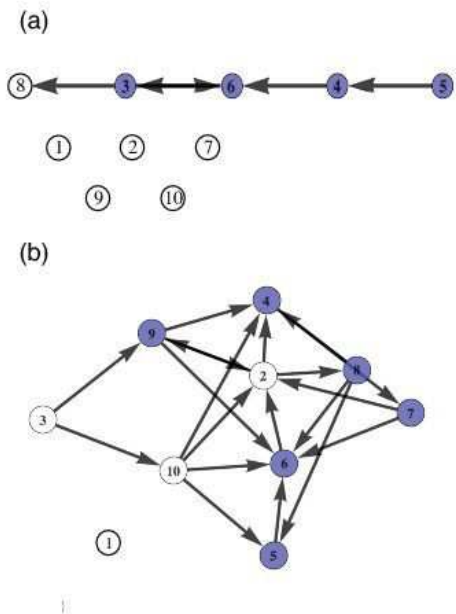} }
\end{center}
\vspace{-15mm}
\caption{(Color online). Two typical realizations of the
synchronization-optimized network at different connectivities (a) $p=0.05$
and (b) $p=0.2$. The blue (gray) nodes indicate entrained oscillators. The numbers in
the nodes are indexes of the oscillators. The parameters are $N=10,%
\protect\beta =\protect\beta _{M}$, $M=11$, and $\protect\delta \protect\beta =10
$.}
\label{fig:typical_graph}
\end{figure}

Typical structures of synchronization-optimized networks are shown in Fig.~%
\ref{fig:typical_graph}. When the connectivity $p$ is small, such networks
usually represent chain fragments. At a higher connectivity, the network
become more complexly organized, as shown in Fig.~\ref{fig:typical_graph}%
(b).

\begin{figure*}[tbp]
\begin{center}
\resizebox{\doublefiguresize}{!}{\includegraphics{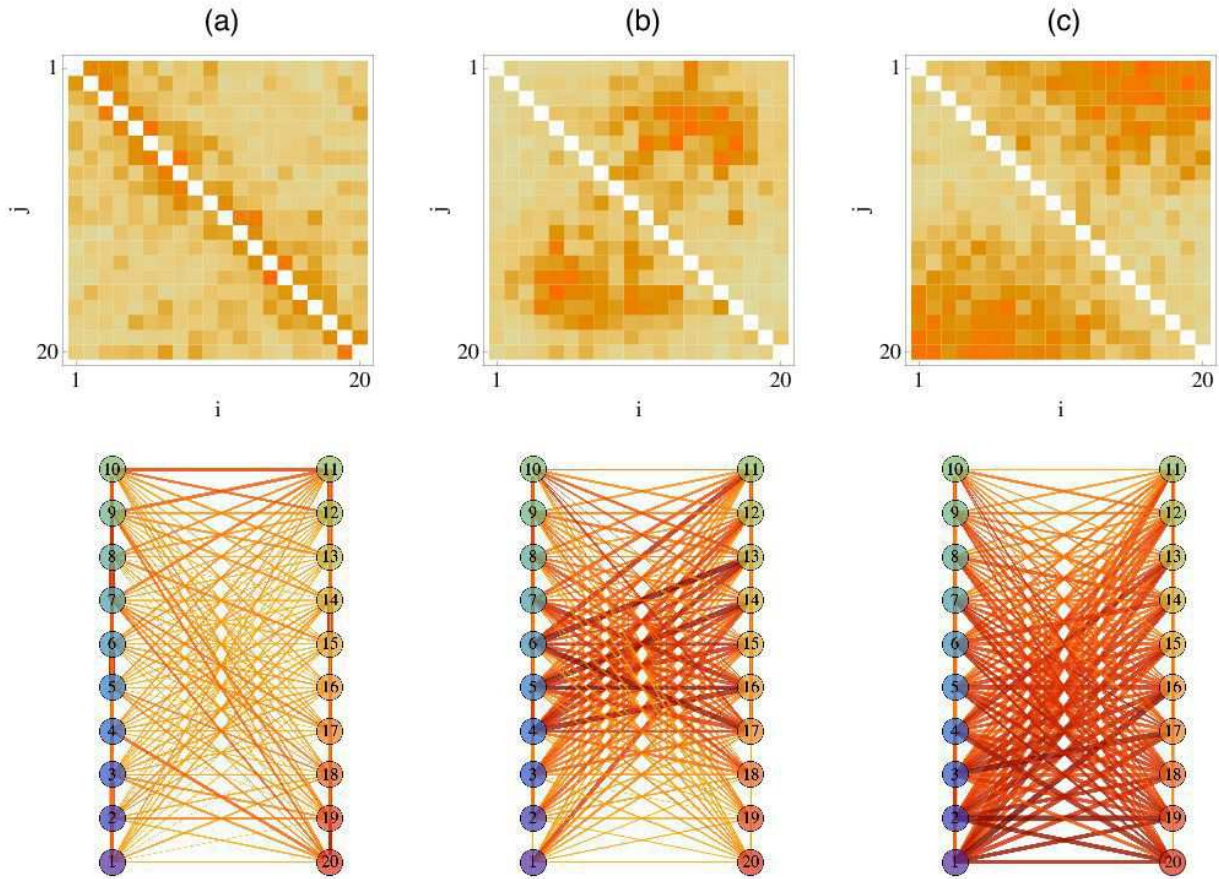} }
\end{center}
\caption{(Color online). The upper panels show adjacent matrices
averaged over the  Gibbs ensemble of synchronization-optimized networks
[see Eq.~(\ref{eq:weightedMatrix})].
The darker color of a matrix element indicates the higher probability of the
respective connections between the elements. The lower panels display the
corresponding network averaged over the Gibbs ensemble. The numbers in the circles show indexes of the oscillators.
The thickness of the lines connecting the nodes
is proportional to the frequency of links between them. The network
connectivity is (a) $p=0.05$, (b) $0.2$ and (c) $0.3$. Other
parameters are same as in Fig.~\protect\ref{fig:timeseries}. }
\label{fig:network1}
\end{figure*}

To statistically characterize the architecture of constructed networks,
ensemble averages of their adjacency matrices over the Gibbs ensemble, i.e.,
\begin{equation}
\overline{\mathbf{a}_{\beta }}=\sum_{\mathbf{a}} \mathbf{a}\exp (\beta R(\mathbf{a}))/Z(\beta ),
\label{eq:weightedMatrix}
\end{equation}
for different connectivities $p$ were computed for $\beta=\beta_M $, as shown in Fig.~\ref%
{fig:network1}. Clearly, the optimal network structure is changing with the
number of links. When the number of links is small, the elements of the mean
adjacency matrix, obtained by averaging over many realization from the
synchronization-optimized ensemble, are large near the diagonal. Hence,
elements with close natural frequency tend to connect and form a chain
fragment. Moreover, oscillators with the natural frequencies near the
center of the interval are often connected. Increasing the number of links,
the network becomes more complicated and off-diagonal elements begin to
dominate instead. The network with the larger $p$ tend to have interlaced
structures, seen in Figs.~\ref{fig:network1}(b)(c), where the oscillators
with roughly opposite natural frequencies are coupled.
A similar trend towards anti-correlations for the oscillators with opposite frequencies has been noticed in  \cite{Brede20085305,Brede08}, where a transition from local to global synchronization under an increase of the coupling strength has been obtained using a different optimization method  \cite{Brede08}. 

This structural transition can be understood as follows: When connectivity
is small, a limited small number of available links is better used to
connect oscillators with frequencies in the middle of the frequency
interval, where the collective synchronization frequency would lie. Indeed,
such oscillators can be easily entrained and even a single link may be
sufficient to synchronize them. If connectivity is increased and some
further links may be used, it would not however be efficient to put them
into the middle region: the oscillators there are already synchronized and
bringing more connections would not increase the performance. This means
that the additional available links should be rather connected to the
elements in the periphery, outside of the central frequency region. If
predominantly local connections between the elements on each side are
established, this would however lead to the formation of two clusters, each
on a different side from the center. Within each cluster, oscillators may
get synchronized, but oscillations of the two clusters will still then be
incoherent. Therefore, a better solution would consist in establishing
pairwise connections between the elements on both sides of the center, i.e.
in linking preferentially the opposite oscillators. This is exactly what we
observe in Fig. \ref{fig:network1} at the higher connectivity $p=0.3$.

\subsection{Degree distributions and cluster organization}

\begin{figure*}[tbp]
\begin{center}
\resizebox{\doublefiguresize}{!}{\includegraphics{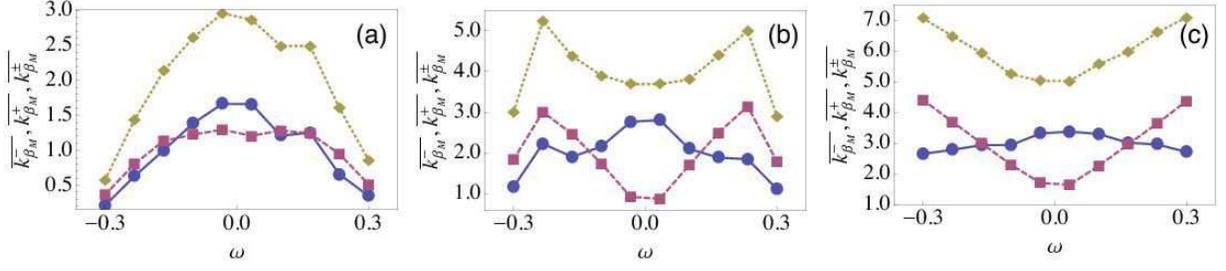} }
\end{center}
\caption{(Color online). Degrees, averaged over the the Gibbs ensemble
of synchronitaion-optimized networks, as  functions of the natural frequency
of the oscillator. The in-degree $\overline{k_{\protect\beta _{M}}^{-}}$,
out-degree $\overline{k_{\protect\beta _{M}}^{+}}$, and degree $\overline{k_{%
\protect\beta _{M}}^{\pm }}$  are plotted by the blue circles, red squares and
yellow diamonds. $N=10,\protect\beta =\protect\beta _{M}$; 
The network connectivity is (a) $p=0.1$,  (b) $p=0.2$ and (c) $p=0.3$.
Other parameters the same as in Fig.~\protect\ref{fig:timeseries}. }
\label{fig:degree}
\end{figure*}

To statistically investigate architectures of designed networks, 
ingoing and outgoing degrees of their nodes have been considered and
averaged over the ensemble. Since the network is colored, i.e, each its node
has a different natural frequency, the mean in- and out-degree of the nodes
can be plotted as a function of their natural frequency (Fig.~\ref{fig:degree}%
). 
When connectivity $p$ is small, both in- and out-degrees averaged over the
ensemble have a maximum at $\omega =0$, i.e, oscillators having smaller
magnitudes of the natural frequency tend to be mutually connected. This
unimodal degree distribution is consistent with the linear chain structure
shown in Fig.~\ref{fig:network1} (a). As $p$ is increased, the mean in-degree
distribution becomes bi-modal and oscillators having larger magnitudes
of the natural frequency tend to have larger out-degrees. This tendency
becomes stronger when $p$ increases [Fig.~\ref{fig:degree} (b)(c)]. 

Furthermore, we calculated the mean numbers of isolated nodes as a function
of $p$. The isolated nodes have been classificed into three categories, as
those which have no in-coming, no out-going, and neither in-coming nor
out-going connections. The numbers of such isolated nodes are, respectively,
\begin{eqnarray}
s^{+}(w) &=&\sum_{i=1}^{N}\Delta (\sum_{j=1}^{N}a_{i,j})  \nonumber \\
s^{-}(w) &=&\sum_{j=1}^{N}\Delta (\sum_{i=1}^{N}a_{i,j})  \nonumber \\
s^{\pm }(w) &=&\sum_{k=1}^{N}\Delta
(\sum_{i=1}^{N}a_{i,k}+\sum_{j=1}^{N}a_{k,j}),
\end{eqnarray}%
where $\Delta (w)$ is the Kronecker symbol, $\Delta (w)=1$ for $w=1$ and $%
\Delta (w)=0$ otherwise. We averaged these numbers over the Gibbs ensemble
for $\beta =\beta _{0}$ and $\beta_{M}$ and determined the ratio $\overline{%
s_{\beta _{M}}^{\pm }}/\overline{s_{\beta _{0}}^{\pm }}$ of the average
number of isolated nodes in the synchronization-optimized networks to that
in the networks obtained by  random rewiring (see Fig.~\ref%
{fig:isolated-node}). 

The results do not depend on the choice of $\beta_m$ qualitatively. 

When $p$ is small, the ratio of completely isolated nodes is larger than
one. This comes from the fact that the links are used intensively between
the nodes having smaller magnitudes of the natural frequency, at the cost of
connections of periphery oscillators. Thus, the number of isolated
nodes is large. Starting from $p\simeq 0.23$, this ratio becomes however
less than one, so that the optimized networks tend to have less completely
isolated nodes as their random counterparts. We can also notice that the
relative number of nodes without ingoing connections becomes high at about $%
p\simeq 0.23$ and then sharply drops down. The number of nodes without the
outgoing connections in the optimized networks remains always larger than in
the random networks.  
\begin{figure}[tbp]
\begin{center}
\resizebox{\singlefiguresize}{!}{\includegraphics{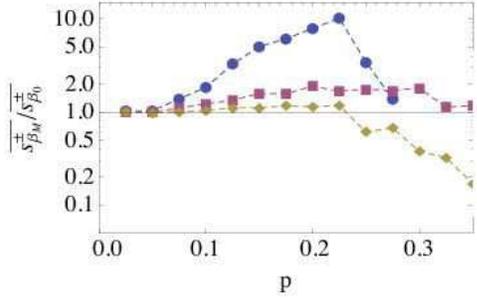} }
\end{center}
\caption{(Color online). The ratio of the number of isolated nodes, averaged
over the replicas with $\protect\beta _{M}$ (synchronization-optimized
networks) to that averaged over the replicas with $\protect\beta _{0}$
(randomly rewired networks) as a function of the connectivity $p$. The data
for the nodes isolated with respect to incoming (blue circles) and outgoing (red squares) connections, as
well as for the completely isolated nodes (yellow diamonds), is shown.  The same parameters as
in Fig.~\protect\ref{fig:timeseries}. }
\label{fig:isolated-node}
\end{figure}

As already suggested by Fig.~\ref{fig:network1} (b)(c),
synchronization-optimized network with larger connectivities may be similar
to bipartite graphs. A bipartite graph is a graph whose nodes can be divided
into two disjoint sets $A$ and $B,$ so that every link connects a node in $A$
to a node in $B$ and \textit{vise versa} \cite{Reinhard05}. To demonstrate
that our optimized networks are indeed similar to bipartite graphs, we
divide all oscillators into two groups $A$ and $B$ with the negative and
positive natural frequencies. An intraconnection is defined as a link
between nodes belonging to the same group, while an interconnection is a
link between the nodes in $A$ and $B$. Thus, the number of intraconnections
is given by 
\[
n^{intra}(w)=(\sum_{i=1,j=N/2}^{N/2,N}+\sum_{i=N/2,j=1}^{N,N/2})a_{i,j},
\]%
and the number of interconnection is  
\[
n^{inter}(w)=(\sum_{i=1,j=1}^{N/2,N/2}+\sum_{i=N/2,j=N/2}^{N,N})a_{i,j}.
\]%
The mean ratio $\overline{n_{\beta _{m}}^{inter}}/\overline{n_{\beta
_{m}}^{intra}}$ of inter- to intraconnections in the synchrony-optimized
ensemble for $\beta_{5},\beta_{10},\beta_{15}$ and $\beta_{20}$ 
as a function of the connectivity $p$ is shown in Fig.~\ref%
{fig:inter-intra}. 
This ratio is smaller than unity when connectivity $p$
is small. It increases with $p$ and reaches a maximum in the vicinity of the
transition point, where the bipartite-like structure emerges. Further above
the transition point, the ratio gradually decreases to unity, since the
number of links increases until all-to-all connections are established 
\footnote{%
Since diagonal elements of the adjacent matrix are chosen to be zero, the
ratio is not equal to one at $p=1$, i.e., for all-to-all connections.}.

\begin{figure}[tbp]
\begin{center}
\resizebox{\singlefiguresize}{!}{\includegraphics{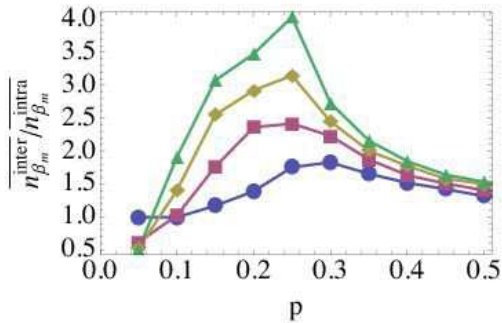}}
\end{center}
\caption{(Color online). The ratio of inter- to intra- connections as a
function of the connectivity $p$ for $\beta=\beta _{5}$ (blue circles), $\beta_{10}$ (red squares), $\beta_{15}$ (yellow diamonds) and $\beta_{20}$ (green triangles).
  $N=10$.
The same other parameters as in Fig.~\protect\ref{fig:timeseries}. }
\label{fig:inter-intra}
\end{figure}

\subsection{Closeness, betweenness and clustering}

To characterize network structure quantitatively, we calculated the closeness, betweenness and clustering
coefficient \cite{Watts98,Boccaletti2006175}. Again, averaging was performed
over many realizations of synchronization-optimized networks, sampled with
the Gibbs distribution (Eq.~\ref{eq:gibbs}).

The betweenness centrality of a node is the number of geodesics (i.e.,
shortest paths) going through it. If there is more than one geodesic between
two nodes, the number of geodesics which connect these two nodes via a
considered node is divided by the total number of geodesics that connect the
two nodes. The betweenness centrality is thus defined by 
\[
C^{btw}(v)=\sum_{s,t}\sum_{s\neq t}\frac{\sigma _{s,t}(v)}{\sigma _{s,t}},
\]%
where $\sigma $ is the number of shortest paths from node $s$ to node $t$
and $\sigma _{s,t}(v)$ is the number of shortest paths from $s$ to $t$ that
pass through node $v$.

The closeness centrality of a node specifies how easily other nodes can be
reached from it (or, in other words, how easily it can be reached from the
other nodes). It is defined as the sum of the lengths of all geodesics
leading to or from the given node, divided by the total number $n$ of nodes
minus one,  
\[
C^{cls}(v)=\frac{1}{n-1}\sum_{t}d_{g}(v,t),
\]%
where $d_{g}(u,v)$ is geodesic distance between the nodes $u$ and $v$ (i.e.,
the length of the shortest path connecting them).

The clustering coefficient of a node specifies the number of neighbours of
this node which are in turn mutual neighbours. It is defined as  
\[
C^{trn}(v)=\frac{t_{v}}{c_{2}^{k_{v}}},
\]%
where $k_{v}$ is the degree of a node $v$ and $t_{v}$ is the number of links
between its neighbors, $c_{2}^{k_{v}}$ is the number of pairs that can be
made by using $k_{v}$ neighbors. 

The above properties are defined for each node. To characterize the entire
network, we average them over all nodes. 
\begin{figure*}[tbp]
\begin{center}
\resizebox{\doublefiguresize}{!}{\includegraphics{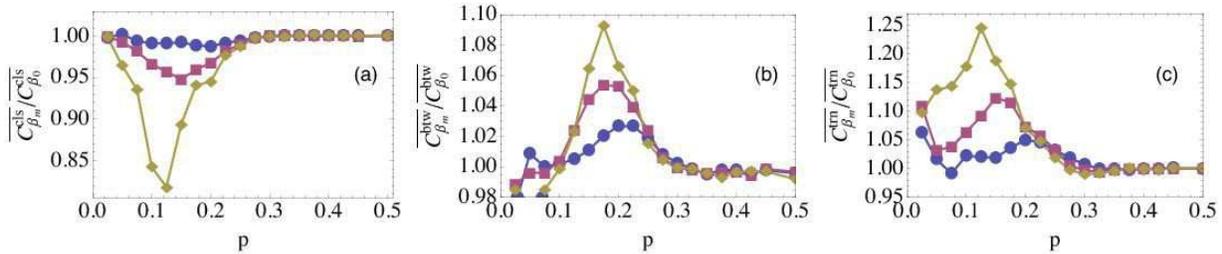} }
\end{center}
\caption{(Color online). Statistical properties of 
synchronization-optimized networks as functions of the connectively $p$.
 (a) the ratio of closeness, averaged over the replicas with $\protect\beta _{m}$ to that averaged over the replicas with $\protect\beta _{0}$, (b) the ratio of betweenness, and (c) the ratio of transitivity, 
 for different $\beta_m$, where $m=10$ (blue circles), $m=15$ (red squares), and $m=20$ (yellow diamonds).   
The same parameters as in Fig.~\protect\ref{fig:timeseries}. }
\label{fig:network_character}
\end{figure*}

In order to quantify differences between synchronization-optimized
networks and networks generated by random rewiring, ratios $\overline{%
C^{K}(w)}_{\beta _{m}}/\overline{C^{K}(w)}_{\beta _{0}}$can be used, where $%
C^{K}$ is the respective property of network, such as closeness,
betweenness, or clustering, $\beta _{m}$ is inverse temperature and $%
\beta _{0}=0$. In Fig.~\ref{fig:network_character}, we show these
ensemble-averaged network properties depending on the connection probability 
$p$ for several inverse temperatures. Obviously, these ratios should approach unity at $p=0$ or at $p=1$,
because the difference in synchronization of optimized and random networks
vanishes in these two limits. The ratios for the closeness have pronounced minimima in the transition region. 
The ratio in the vicinity of the transition point decreases when the performance of optimized network increases, i.e., the network ensemble with higher inverse temperature.

On the other hand, the betweenness and
clustering coefficient gradually increase with the connectivity $p$ and
reach a maximum in the transition region. Note that in recent work \cite{PhysRevE.72.015101} it was found that, both in random and scale-free networks, increase the clustering coefficient
favors formation of oscillator sub-populations synchronized at different
frequencies.

\section{Conclusions}
~\label{sec:summary}
We have designed synchronization-optimized networks with a fixed number of
links for a heterogeneous oscillator population. This has been done by using
the Markov Chain Stochastic Monte Carlo method complemented by the Replica
Exchange algorithm. A transition from the linear to bipartite-like networks
has been found under increasing the number of links. At low connectivity,
synchronization-optimized networks typically represent small chains
connecting oscillators with close natural frequencies. As the number of
links increases, the networks become interlaced and oscillators with
opposite natural frequencies tend to be connected. Therefore,
synchronization-optimized network begin to resemble bipartite graphs. This
structural change of synchronion-optimized network is clearly revealed
through the analysis of inter- and intraconnections.

Thus, we have shown that the efficient design of oscillator networks with
the improved synchronization properties is possible. The architectures of
such optimal networks strongly depend on the constraints, such as the total
number of links available. Through the appropriate rewiring of a network, a
strong gain in the synchronization signal can be achieved. 

Although our study has been performed for a simple system of phase
oscillators, similar evolutionary optimization methods can be applied to
construct networks of different origins,
where the dynamics of individual oscillators may be significantly more
complex. 

\section*{Acknowledgments}

This study has been partially supported by the Ministry of Education,
Science, Sports and Culture, Grant-in-Aid for Scientific Research (21540376)
and the Volkswagen Foundation (Germany).


\section*{References}

\bibliography{refkuramoto,refmcmc}

\begin{thebibliography}{39}
\expandafter\ifx\csname natexlab\endcsname\relax\def\natexlab#1{#1}\fi
\expandafter\ifx\csname bibnamefont\endcsname\relax
  \def\bibnamefont#1{#1}\fi
\expandafter\ifx\csname bibfnamefont\endcsname\relax
  \def\bibfnamefont#1{#1}\fi
\expandafter\ifx\csname citenamefont\endcsname\relax
  \def\citenamefont#1{#1}\fi
\expandafter\ifx\csname url\endcsname\relax
  \def\url#1{\texttt{#1}}\fi
\expandafter\ifx\csname urlprefix\endcsname\relax\def\urlprefix{URL }\fi
\providecommand{\bibinfo}[2]{#2}
\providecommand{\eprint}[2][]{\url{#2}}

\bibitem[{\citenamefont{Albert and Barab\'asi}(2002)}]{RevModPhys.74.47}
\bibinfo{author}{\bibfnamefont{R.}~\bibnamefont{Albert}} \bibnamefont{and}
  \bibinfo{author}{\bibfnamefont{A.-L.} \bibnamefont{Barab\'asi}},
  \bibinfo{journal}{Rev. Mod. Phys.} \textbf{\bibinfo{volume}{74}},
  \bibinfo{pages}{47} (\bibinfo{year}{2002}).

\bibitem[{\citenamefont{Boccaletti et~al.}(2006)\citenamefont{Boccaletti,
  Latora, Moreno, Chavez, and Hwang}}]{Boccaletti2006175}
\bibinfo{author}{\bibfnamefont{S.}~\bibnamefont{Boccaletti}},
  \bibinfo{author}{\bibfnamefont{V.}~\bibnamefont{Latora}},
  \bibinfo{author}{\bibfnamefont{Y.}~\bibnamefont{Moreno}},
  \bibinfo{author}{\bibfnamefont{M.}~\bibnamefont{Chavez}}, \bibnamefont{and}
  \bibinfo{author}{\bibfnamefont{D.-U.} \bibnamefont{Hwang}},
  \bibinfo{journal}{Phys. Rep.} \textbf{\bibinfo{volume}{424}},
  \bibinfo{pages}{175 } (\bibinfo{year}{2006}).

\bibitem[{\citenamefont{Manrubia et~al.}(2004)\citenamefont{Manrubia,
  Mikhailov, and Zanette}}]{Manrubia04}
\bibinfo{author}{\bibfnamefont{S.}~\bibnamefont{Manrubia}},
  \bibinfo{author}{\bibfnamefont{A.}~\bibnamefont{Mikhailov}},
  \bibnamefont{and} \bibinfo{author}{\bibfnamefont{D.}~\bibnamefont{Zanette}},
  \emph{\bibinfo{title}{Emergence of Dynamical Order: Synchronization Phenomena
  in Complex Systems}} (\bibinfo{publisher}{World Scientific, Singapore},
  \bibinfo{year}{2004}).

\bibitem[{\citenamefont{Arenas et~al.}(2008)\citenamefont{Arenas, Diaz-Guilera,
  Kurths, Moreno, and Zhou}}]{Arenas08}
\bibinfo{author}{\bibfnamefont{A.}~\bibnamefont{Arenas}},
  \bibinfo{author}{\bibfnamefont{A.}~\bibnamefont{Diaz-Guilera}},
  \bibinfo{author}{\bibfnamefont{J.}~\bibnamefont{Kurths}},
  \bibinfo{author}{\bibfnamefont{Y.}~\bibnamefont{Moreno}}, \bibnamefont{and}
  \bibinfo{author}{\bibfnamefont{C.}~\bibnamefont{Zhou}},
  \bibinfo{journal}{Phys. Rep.} \textbf{\bibinfo{volume}{496}},
  \bibinfo{pages}{93} (\bibinfo{year}{2008}).

\bibitem[{\citenamefont{Kurths et~al.}(2001)\citenamefont{Kurths, Pikovsky, and
  Rosenblum}}]{Kurths01}
\bibinfo{author}{\bibfnamefont{J.}~\bibnamefont{Kurths}},
  \bibinfo{author}{\bibfnamefont{A.}~\bibnamefont{Pikovsky}}, \bibnamefont{and}
  \bibinfo{author}{\bibfnamefont{M.}~\bibnamefont{Rosenblum}},
  \emph{\bibinfo{title}{Synchronization: A Universal Concept in Nonlinear
  Sciences}} (\bibinfo{publisher}{Cambridge Univ. Press, Cambridge},
  \bibinfo{year}{2001}).

\bibitem[{\citenamefont{Hong et~al.}(2002)\citenamefont{Hong, Choi, and
  Kim}}]{Hong02}
\bibinfo{author}{\bibfnamefont{H.}~\bibnamefont{Hong}},
  \bibinfo{author}{\bibfnamefont{M.~Y.} \bibnamefont{Choi}}, \bibnamefont{and}
  \bibinfo{author}{\bibfnamefont{B.~J.} \bibnamefont{Kim}},
  \bibinfo{journal}{Phys. Rev. E} \textbf{\bibinfo{volume}{65}},
  \bibinfo{pages}{026139} (\bibinfo{year}{2002}).

\bibitem[{\citenamefont{Ichinomiya}(2004)}]{Ichinomiya04}
\bibinfo{author}{\bibfnamefont{T.}~\bibnamefont{Ichinomiya}},
  \bibinfo{journal}{Phys. Rev. E} \textbf{\bibinfo{volume}{70}},
  \bibinfo{pages}{026116} (\bibinfo{year}{2004}).

\bibitem[{\citenamefont{Lee}(2005)}]{Lee05}
\bibinfo{author}{\bibfnamefont{D.-S.} \bibnamefont{Lee}},
  \bibinfo{journal}{Phys. Rev. E} \textbf{\bibinfo{volume}{72}},
  \bibinfo{pages}{026208} (\bibinfo{year}{2005}).

\bibitem[{\citenamefont{Motter et~al.}(2005)\citenamefont{Motter, Zhou, and
  Kurths}}]{PhysRevE.71.016116}
\bibinfo{author}{\bibfnamefont{A.~E.} \bibnamefont{Motter}},
  \bibinfo{author}{\bibfnamefont{C.}~\bibnamefont{Zhou}}, \bibnamefont{and}
  \bibinfo{author}{\bibfnamefont{J.}~\bibnamefont{Kurths}},
  \bibinfo{journal}{Phys. Rev. E} \textbf{\bibinfo{volume}{71}},
  \bibinfo{pages}{016116} (\bibinfo{year}{2005}).

\bibitem[{\citenamefont{Hwang et~al.}(2005)\citenamefont{Hwang, Chavez, Amann,
  and Boccaletti}}]{PhysRevLett.94.138701}
\bibinfo{author}{\bibfnamefont{D.-U.} \bibnamefont{Hwang}},
  \bibinfo{author}{\bibfnamefont{M.}~\bibnamefont{Chavez}},
  \bibinfo{author}{\bibfnamefont{A.}~\bibnamefont{Amann}}, \bibnamefont{and}
  \bibinfo{author}{\bibfnamefont{S.}~\bibnamefont{Boccaletti}},
  \bibinfo{journal}{Phys. Rev. Lett.} \textbf{\bibinfo{volume}{94}},
  \bibinfo{pages}{138701} (\bibinfo{year}{2005}).

\bibitem[{\citenamefont{Ipsen and Mikhailov}(2002)}]{Mikhailov02}
\bibinfo{author}{\bibfnamefont{M.}~\bibnamefont{Ipsen}} \bibnamefont{and}
  \bibinfo{author}{\bibfnamefont{A.~S.} \bibnamefont{Mikhailov}},
  \bibinfo{journal}{Phys. Rev. E} \textbf{\bibinfo{volume}{66}},
  \bibinfo{pages}{046109} (\bibinfo{year}{2002}).

\bibitem[{\citenamefont{{Moyano} et~al.}(2001)\citenamefont{{Moyano},
  {Abramson}, and {Zanette}}}]{Moyano01}
\bibinfo{author}{\bibfnamefont{L.~G.} \bibnamefont{{Moyano}}},
  \bibinfo{author}{\bibfnamefont{G.}~\bibnamefont{{Abramson}}},
  \bibnamefont{and} \bibinfo{author}{\bibfnamefont{D.~H.}
  \bibnamefont{{Zanette}}}, \bibinfo{journal}{Eur. Phys. J. B}
  \textbf{\bibinfo{volume}{22}}, \bibinfo{pages}{223} (\bibinfo{year}{2001}).

\bibitem[{\citenamefont{Gleiser and Zanette}(2006)}]{Gleiser06}
\bibinfo{author}{\bibfnamefont{P.~M.} \bibnamefont{Gleiser}} \bibnamefont{and}
  \bibinfo{author}{\bibfnamefont{D.~H.} \bibnamefont{Zanette}},
  \bibinfo{journal}{Eur. Phys. J. B} \textbf{\bibinfo{volume}{53}},
  \bibinfo{pages}{233} (\bibinfo{year}{2006}).

\bibitem[{\citenamefont{Buzna et~al.}(2009)\citenamefont{Buzna, Lozano, and
  D\'\i{}az-Guilera}}]{PhysRevE.80.066120}
\bibinfo{author}{\bibfnamefont{L.}~\bibnamefont{Buzna}},
  \bibinfo{author}{\bibfnamefont{S.}~\bibnamefont{Lozano}}, \bibnamefont{and}
  \bibinfo{author}{\bibfnamefont{A.}~\bibnamefont{D\'\i{}az-Guilera}},
  \bibinfo{journal}{Phys. Rev. E} \textbf{\bibinfo{volume}{80}},
  \bibinfo{pages}{066120} (\bibinfo{year}{2009}).

\bibitem[{\citenamefont{Brede}(2008{\natexlab{a}})}]{Brede20085305}
\bibinfo{author}{\bibfnamefont{M.}~\bibnamefont{Brede}},
  \bibinfo{journal}{Phys. Lett. A} \textbf{\bibinfo{volume}{372}},
  \bibinfo{pages}{5305 } (\bibinfo{year}{2008}{\natexlab{a}}), ISSN
  \bibinfo{issn}{0375-9601}.

\bibitem[{\citenamefont{Brede}(2008{\natexlab{b}})}]{Brede08}
\bibinfo{author}{\bibfnamefont{M.}~\bibnamefont{Brede}}, \bibinfo{journal}{Eur.
  Phys. J. B} \textbf{\bibinfo{volume}{62}}, \bibinfo{pages}{87}
  (\bibinfo{year}{2008}{\natexlab{b}}).

\bibitem[{\citenamefont{Kuramoto}(1984)}]{Kuramoto84}
\bibinfo{author}{\bibfnamefont{Y.}~\bibnamefont{Kuramoto}},
  \emph{\bibinfo{title}{Chemical Oscillations, Waves, and Turbulence}}
  (\bibinfo{publisher}{Springer}, \bibinfo{year}{1984}).

\bibitem[{\citenamefont{Acebr\'on et~al.}(2005)\citenamefont{Acebr\'on,
  Bonilla, P\'erez~Vicente, Ritort, and Spigler}}]{RevModPhys77137}
\bibinfo{author}{\bibfnamefont{J.~A.} \bibnamefont{Acebr\'on}},
  \bibinfo{author}{\bibfnamefont{L.~L.} \bibnamefont{Bonilla}},
  \bibinfo{author}{\bibfnamefont{C.~J.} \bibnamefont{P\'erez~Vicente}},
  \bibinfo{author}{\bibfnamefont{F.}~\bibnamefont{Ritort}}, \bibnamefont{and}
  \bibinfo{author}{\bibfnamefont{R.}~\bibnamefont{Spigler}},
  \bibinfo{journal}{Rev. Mod. Phys.} \textbf{\bibinfo{volume}{77}},
  \bibinfo{pages}{137} (\bibinfo{year}{2005}).

\bibitem[{\citenamefont{Tanaka and Aoyagi}(2008)}]{tanaka08}
\bibinfo{author}{\bibfnamefont{T.}~\bibnamefont{Tanaka}} \bibnamefont{and}
  \bibinfo{author}{\bibfnamefont{T.}~\bibnamefont{Aoyagi}},
  \bibinfo{journal}{Phys. Rev. E} \textbf{\bibinfo{volume}{78}},
  \bibinfo{eid}{046210} (\bibinfo{year}{2008}).

\bibitem[{\citenamefont{Gong and van Leeuwen}(2004)}]{Gong04}
\bibinfo{author}{\bibfnamefont{P.}~\bibnamefont{Gong}} \bibnamefont{and}
  \bibinfo{author}{\bibfnamefont{C.}~\bibnamefont{van Leeuwen}},
  \bibinfo{journal}{Europhys. Lett.} \textbf{\bibinfo{volume}{67}},
  \bibinfo{pages}{328} (\bibinfo{year}{2004}).

\bibitem[{\citenamefont{van~den Berg and van Leeuwen}(2004)}]{Berg04}
\bibinfo{author}{\bibfnamefont{D.}~\bibnamefont{van~den Berg}}
  \bibnamefont{and} \bibinfo{author}{\bibfnamefont{C.}~\bibnamefont{van
  Leeuwen}}, \bibinfo{journal}{Europhys. Lett.} \textbf{\bibinfo{volume}{65}},
  \bibinfo{pages}{459} (\bibinfo{year}{2004}).

\bibitem[{\citenamefont{Landau and Binder}(2005)}]{Landau05}
\bibinfo{author}{\bibfnamefont{D.}~\bibnamefont{Landau}} \bibnamefont{and}
  \bibinfo{author}{\bibfnamefont{K.}~\bibnamefont{Binder}},
  \emph{\bibinfo{title}{A Guide to Monte Carlo Simulations in Statistical
  Physics}} (\bibinfo{publisher}{Cambridge University Press},
  \bibinfo{year}{2005}).

\bibitem[{\citenamefont{Newman and Barkema}(1999)}]{Newman99}
\bibinfo{author}{\bibfnamefont{M.~E.~J.} \bibnamefont{Newman}}
  \bibnamefont{and} \bibinfo{author}{\bibfnamefont{G.~T.}
  \bibnamefont{Barkema}}, \emph{\bibinfo{title}{Monte Carlo Methods in
  Statistical Physics}} (\bibinfo{publisher}{Oxford University Press},
  \bibinfo{year}{1999}).

\bibitem[{\citenamefont{Liu}(2001)}]{Liu01}
\bibinfo{author}{\bibfnamefont{J.}~\bibnamefont{Liu}},
  \emph{\bibinfo{title}{Monte Carlo Strategies in Scientific Computing}}
  (\bibinfo{publisher}{Springer}, \bibinfo{year}{2001}).

\bibitem[{\citenamefont{Cho et~al.}(1994)\citenamefont{Cho, Doll, and
  Freeman}}]{Cho94}
\bibinfo{author}{\bibfnamefont{A.~E.} \bibnamefont{Cho}},
  \bibinfo{author}{\bibfnamefont{J.~D.} \bibnamefont{Doll}}, \bibnamefont{and}
  \bibinfo{author}{\bibfnamefont{D.~L.} \bibnamefont{Freeman}},
  \bibinfo{journal}{Chem. Phys. Lett.} \textbf{\bibinfo{volume}{229}},
  \bibinfo{pages}{218} (\bibinfo{year}{1994}).

\bibitem[{\citenamefont{Bolhuis et~al.}(1998)\citenamefont{Bolhuis, Dellago,
  and Chandler}}]{Bolhuis98}
\bibinfo{author}{\bibfnamefont{P.~G.} \bibnamefont{Bolhuis}},
  \bibinfo{author}{\bibfnamefont{C.}~\bibnamefont{Dellago}}, \bibnamefont{and}
  \bibinfo{author}{\bibfnamefont{D.}~\bibnamefont{Chandler}},
  \bibinfo{journal}{Faraday Discuss.} \textbf{\bibinfo{volume}{110}},
  \bibinfo{pages}{421} (\bibinfo{year}{1998}).

\bibitem[{\citenamefont{Vlugt and Smit}(2000)}]{Vlugt00}
\bibinfo{author}{\bibfnamefont{T.}~\bibnamefont{Vlugt}} \bibnamefont{and}
  \bibinfo{author}{\bibfnamefont{B.}~\bibnamefont{Smit}},
  \bibinfo{journal}{Phys. Chem. Comm.} \textbf{\bibinfo{volume}{2}},
  \bibinfo{pages}{Art. No. 2} (\bibinfo{year}{2000}).

\bibitem[{\citenamefont{Kawasaki and Sasa}(2005)}]{Kawasaki05}
\bibinfo{author}{\bibfnamefont{M.}~\bibnamefont{Kawasaki}} \bibnamefont{and}
  \bibinfo{author}{\bibfnamefont{S.~I.} \bibnamefont{Sasa}},
  \bibinfo{journal}{Phys. Rev. E} \textbf{\bibinfo{volume}{72}},
  \bibinfo{pages}{037202} (\bibinfo{year}{2005}).

\bibitem[{\citenamefont{Sasa and Hayashi}(2006)}]{Sasa06}
\bibinfo{author}{\bibfnamefont{S.~I.} \bibnamefont{Sasa}} \bibnamefont{and}
  \bibinfo{author}{\bibfnamefont{K.}~\bibnamefont{Hayashi}},
  \bibinfo{journal}{Europhys. Lett.} \textbf{\bibinfo{volume}{76}},
  \bibinfo{pages}{156} (\bibinfo{year}{2006}).

\bibitem[{\citenamefont{Giardin\'{a} et~al.}(2006)\citenamefont{Giardin\'{a},
  Kurchan, and Peliti}}]{Giardin06}
\bibinfo{author}{\bibfnamefont{C.}~\bibnamefont{Giardin\'{a}}},
  \bibinfo{author}{\bibfnamefont{J.}~\bibnamefont{Kurchan}}, \bibnamefont{and}
  \bibinfo{author}{\bibfnamefont{L.}~\bibnamefont{Peliti}},
  \bibinfo{journal}{Phys. Rev. Lett.} \textbf{\bibinfo{volume}{96}},
  \bibinfo{pages}{120603} (\bibinfo{year}{2006}).

\bibitem[{\citenamefont{Tailleur and Kurchan}(2007)}]{Tailleur07}
\bibinfo{author}{\bibfnamefont{J.}~\bibnamefont{Tailleur}} \bibnamefont{and}
  \bibinfo{author}{\bibfnamefont{J.}~\bibnamefont{Kurchan}},
  \bibinfo{journal}{Nature Physics 3} \textbf{\bibinfo{volume}{3}},
  \bibinfo{pages}{203} (\bibinfo{year}{2007}).

\bibitem[{\citenamefont{Yanagita and Iba}(2009)}]{yanagita09}
\bibinfo{author}{\bibfnamefont{T.}~\bibnamefont{Yanagita}} \bibnamefont{and}
  \bibinfo{author}{\bibfnamefont{Y.}~\bibnamefont{Iba}}, \bibinfo{journal}{J.
  Stat. Mech.} \textbf{\bibinfo{volume}{2}}, \bibinfo{pages}{02043}
  (\bibinfo{year}{2009}).

\bibitem[{\citenamefont{Metropolis et~al.}(1953)\citenamefont{Metropolis,
  Rosenbluth, Rosenbluth, Teller, and Teller}}]{Metropolis53}
\bibinfo{author}{\bibfnamefont{N.}~\bibnamefont{Metropolis}},
  \bibinfo{author}{\bibfnamefont{A.}~\bibnamefont{Rosenbluth}},
  \bibinfo{author}{\bibfnamefont{M.}~\bibnamefont{Rosenbluth}},
  \bibinfo{author}{\bibfnamefont{A.}~\bibnamefont{Teller}}, \bibnamefont{and}
  \bibinfo{author}{\bibfnamefont{E.}~\bibnamefont{Teller}},
  \bibinfo{journal}{J. Chem. Phys.} \textbf{\bibinfo{volume}{21}},
  \bibinfo{pages}{1087} (\bibinfo{year}{1953}).

\bibitem[{\citenamefont{Hukushima and Nemoto}(1996)}]{Hukushima96}
\bibinfo{author}{\bibfnamefont{K.}~\bibnamefont{Hukushima}} \bibnamefont{and}
  \bibinfo{author}{\bibfnamefont{K.}~\bibnamefont{Nemoto}},
  \bibinfo{journal}{J. Phys. Soc. Jpn.} \textbf{\bibinfo{volume}{65}},
  \bibinfo{pages}{1604} (\bibinfo{year}{1996}).

\bibitem[{\citenamefont{Iba}(2001)}]{Iba01}
\bibinfo{author}{\bibfnamefont{Y.}~\bibnamefont{Iba}}, \bibinfo{journal}{Int.
  J. Mod. Phys. C} \textbf{\bibinfo{volume}{12}}, \bibinfo{pages}{623}
  (\bibinfo{year}{2001}).

\bibitem[{\citenamefont{Janke}(2008)}]{Janke08}
\bibinfo{editor}{\bibfnamefont{W.}~\bibnamefont{Janke}}, ed.,
  \emph{\bibinfo{title}{Rugged Free Energy Landscapes: Common Computational
  Approaches to Spin Glasses, Structural Glasses and Biological
  Macromolecules}}, Lect. Notes Phys. Vol. 736 (\bibinfo{publisher}{Springer,
  Berlin}, \bibinfo{year}{2008}).

\bibitem[{\citenamefont{Diestel}(2005)}]{Reinhard05}
\bibinfo{author}{\bibfnamefont{R.}~\bibnamefont{Diestel}},
  \emph{\bibinfo{title}{Graph Theory}} (\bibinfo{publisher}{Springer},
  \bibinfo{year}{2005}).

\bibitem[{\citenamefont{Watts and Strogatz}(1998)}]{Watts98}
\bibinfo{author}{\bibfnamefont{D.~J.} \bibnamefont{Watts}} \bibnamefont{and}
  \bibinfo{author}{\bibfnamefont{S.~H.} \bibnamefont{Strogatz}},
  \bibinfo{journal}{Nature} \textbf{\bibinfo{volume}{393}}, \bibinfo{pages}{440
  } (\bibinfo{year}{1998}).

\bibitem[{\citenamefont{McGraw and Menzinger}(2005)}]{PhysRevE.72.015101}
\bibinfo{author}{\bibfnamefont{P.~N.} \bibnamefont{McGraw}} \bibnamefont{and}
  \bibinfo{author}{\bibfnamefont{M.}~\bibnamefont{Menzinger}},
  \bibinfo{journal}{Phys. Rev. E} \textbf{\bibinfo{volume}{72}},
  \bibinfo{pages}{015101(R)} (\bibinfo{year}{2005}).

\end{thebibliography}

\end{document}